\documentclass{llncs}

\usepackage{algorithmic}
\usepackage{algorithm}
\usepackage{amsmath}
\usepackage{comment}
\usepackage{fancybox}
\usepackage{subfigure}
\usepackage{tabularx}
\usepackage{url}
\usepackage{color}
\usepackage{graphicx}
\usepackage{listings}
\usepackage{syntax}
\usepackage{wrapfig}
\title{Data Poisoning: Lightweight Soft Fault Injection for Python}
\author{Mohammad Amin Alipour, Alex Groce
\institute{Oregon State University}
}

\newcommand{\prog}{\texttt{Prog}}
\newcommand{\SSA}{\texttt{SSA}}
\newcommand{\poi}[1]{$\partial(#1)$}

\newcommand{\Fix}[1]{\textcolor{red}{[#1]}}

\newtheorem{defin}{Definition}

\begin{document}
\maketitle

\lstset{language=Python,
  basicstyle=\scriptsize\ttfamily,        
  keywordstyle=\color{blue}       
}          

\begin{abstract}
This paper introduces and explores the idea of data poisoning, 
a light-weight peer-architecture technique to inject faults into 
Python programs. This method requires very small modification to
the original program, which facilitates evaluation of sensitivity 
of systems that are prototyped or modeled in Python.
We propose different fault scenarios that can be injected to 
programs using data poisoning. 
We use  Dijkstra's Self Stabilizing Ring Algorithm to illustrate 
the approach.
\end{abstract}

\section{Introduction}
Designing dependable software is notoriously hard. 
In designing critical systems, designers should continuously evaluate the
dependability of their design in the presence of potential faults or failures.
Fault injection tools emulate potential problems in the environment that 
a system will operate within \cite{hsueh1997fault}, by injecting an
error state into the running program.   We note that given the
traditional meanings of fault, error, and failure in software testing,
error injection is possibly a better name for this method.

Broadly speaking, there are two primary fault injection techniques:
hardware-based fault injection and software-based fault injection. In
the first, additional hardware is used to inject faults. This hardware
introduces an error via contact with a target circuit by changing
voltage or current, or uses radiation or some other physical
phenomena to introduce a fault.

Software-based fault injection approaches use code instrumentation or
runtime injection.  In the code instrumentation approach, additional
code is inserted into source code to emulate faults. In runtime
injection, events in the computational environment such as interrupts
and timeouts are used to introduce the fault.  In this paper, we
introduce a software-based approach, called \emph{data poisoning}, for fault injection exploiting
dynamism in the dispatching mechanism of the Python programming
language for fault injection.

Python is a popular programming language for fast prototyping.  There
are growing number of libraries in Python that help developers to
prototype or model system in Python: e.g., MyHDL for
hardware design, SimPy for discrete event simulation, Kairos
\cite{kairos} for programming sensor networks, and Pymote for simulation
of distributed algorithms.

The Python data model abstracts all data as objects \cite{datamodel}. It also simplifies
making changes in the basic behavior of objects. This flexible data model
has been utilized to create dynamic symbolic execution engines for
Python \cite{seer,ball2015}, in which
concrete objects in the program are replaced with  proxy objects that carry out
symbolic execution in addition to concrete computation. 

\emph{Data poisoning} exploits the flexibility of the Python data
model to inject faults with minimal source code changes and a
degree of control over how faults and injected and propagated that
greatly  exceeds that of traditional fault injection tools.  The
contributions of this short paper are:  (1) a definition of data
poisoning and (2) an implementation of data
poisoning for Python.

\section{Data Poisoning}
In this section, we define data poisoning and provide some examples to
clarify the core concepts. 

Table \ref{tbl:notation} shows the notation that we use in the remained of 
this paper. Suppose a program \prog{} in \SSA{} form. That is,  all variables in 
\prog{} are assigned to only once.
Predicate $uses(S,v)$ denotes whether statement $S$ uses variable $v$. 
For example, if $s=$ \texttt{velocity = v0 + acceleration*t} and $v=$\texttt{acceleration},
predicate $uses(s,v)$ holds, because \texttt{acceleration} is used in $S$.

$dev(S,v)$ denotes deviation in the behavior of an operator of a variable $v$ in statement $S$.
For example, if the result of multiplication \texttt{*} in $s$ is 1\% more than the correct
value, $dev(s, acceleration)$ holds. Note that $dev(S,v) \implies uses(S,v)$.

\begin{defin}[Data Poisoning]
Variable $v$ is poisoned, or \poi{v},  if and only if there is a statement $s$ in \prog{} such that if $s$ is 
executed long enough the probability of $dev(s,v)$ approaches 1.
\end{defin}

\begin{table}[t]
\centering
\caption{Notation Guide}
\label{tbl:notation}
\begin{tabular}{|c|l|} 
\hline
Notation & Meaning \\ \hline \hline
\poi{v}      &  \texttt{true},  if $v$ is poisoned, \\
        &  \texttt{false}, otherwise \\ \hline
$uses(S, v)$ & \texttt{true}, if statement $S$ uses $v$ \\
        &  \texttt{false}, otherwise \\ \hline

$dev(S,v)$ & \texttt{true}, if  poisoned $v$ that is used in $S$ has manifested deviant behavior\\
     &  \texttt{false}, otherwise \\ \hline
\end{tabular}
\end{table}

\section{Implementation}
In this section, we describe the architecture for data poisoning. We also use an example 
to illustrate data poisoning at work. 

\begin{wrapfigure}{r}{0.4\textwidth}
\centering
\includegraphics[scale=0.4]{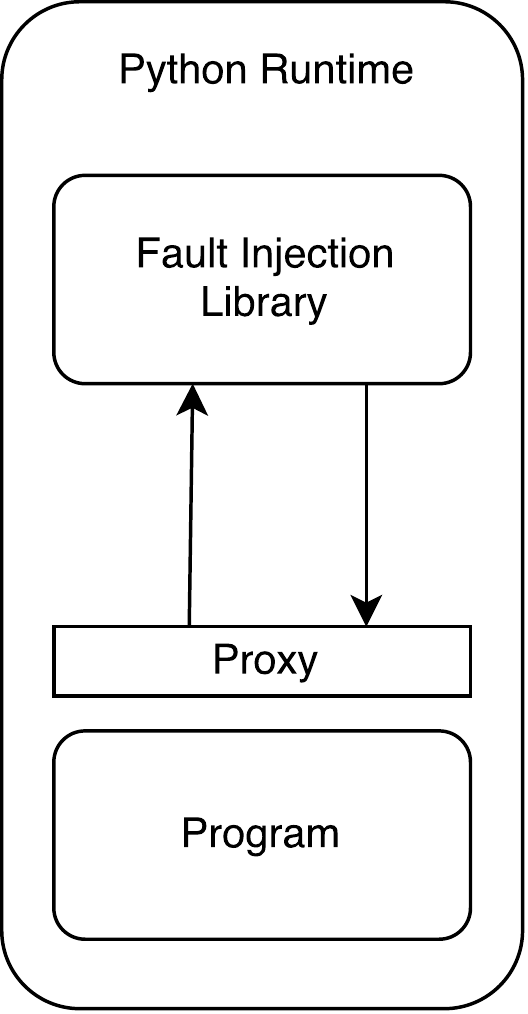}
\caption{Architecture}
\label{fig:architecture}
\end{wrapfigure}

Figure \ref{fig:architecture} illustrates the peer architecture for data poisoning.
Target objects are replaced with corresponding  poisoned proxy objects. 
Responsibility of proxy objects is to receive program's operator callbacks and 
depending on fault injection configurations process them.

\subsection{Example}
Self-stabilizing systems, when perturbed to an error
state, are able to converge to a non-error state in a finite number of steps. 
Such systems are of particular interest in distributed fault-tolerant systems.
Designing such systems is not trivial and using fault injection can uncover
problems in designs before designers attempt to prove correctness of the design 
\cite{perner2013byzantine}.

Dijkstra assumes a ring network with $N+1$ nodes numbered as $0, ..,
N$.  Each node can be in one of $K$ states, where $K>N$. In a non-error
state only one node has the privilege to transition to a new
state. The privilege can be seen as a token in the system for
performing an action, and there must be only one token at a time in
the system; furthermore, all nodes must eventually get the privilege.
Making this system self-stabilizing means that if at any time there
are $0$ or $> 1$ tokens in the system, they system should
eventually move to the state with only one token in the
system.  A fault injection tool and rapid prototyping can help
designers develop and refine such algorithms.

Suppose $S$ shows the current state of a node $i$ and $L$ is the
current state of its left node,
Dijkstra's algorithm for this problem can be described as below:
\begin{align*}
n = 0 \wedge L=S &\implies S = (S+1)\%K \\
n \neq 0 \wedge L\neq S &\implies S = L 
\end{align*}
The key to understanding this algorithm is to observe the behavior of the algorithm when
the state of system is perturbed to an error state.

Figure \ref{code:sss} shows snippets of implementation of Dijkstra's
algorithm.  The 
\texttt{Node} class represents nodes. \texttt{status}  shows the current
status of an instance class of \texttt{Node}. It is initialized to \texttt{0}. Function 
\texttt{update} updates the status of the node based on the algorithm.
In \texttt{update}, whenever a node can make a transition to new state, it obtains
the privilege. To observe the state of the system, it calls function \texttt{out()} 
from \texttt{Network} class. \texttt{Network} class contains a list of nodes in the network.
\texttt{out} returns a string that represent the state of all nodes in the network. 
It gets the state of each node, by calling its \texttt{MON\_\_hasPrivilege} function.
Our tool considers the functions that their names start with \texttt{MON\_\_} as 
the monitoring/reporting functions and  disables the data poisoning in them.

\begin{figure}[t]
{\scriptsize
\begin{lstlisting}
class Node:
  nodecount = 0
  def __init__(self):
    self.status = 0 # Poison(0, size=1, rate=1, cascade=True)
    self.nid = Node.nodecount
    Node.nodecount += 1
    self.privilege = False

  def MON__hasPrivilege(self):
    L = self.getLeft()
    S = self.status
    if self.nid == 0:
      return L == S
    else:
      return L != S

  def update(self):
    L = self.getLeft()
    S = self.status
    self.privilege = False
    if self.nid == 0 and L == S:
      self.privilege = True
      print(Network.out())
      self.privilege = False
      self.status = (self.status + 1) % K
    if self.nid != 0 and L != self.status:
      self.privilege = True
      print(Network.out())
      self.privilege = False
      self.status = L
\end{lstlisting}
}
\caption{Dijkstra's Self-Stabilizing Algorithm}
\label{code:sss}
\end{figure}

Given the definition of class \texttt{Node}, one can write a small simulator as 
in Figure \ref{code:simulator}. This simulator creates a network of size \texttt{5}
and \texttt{10} times updates the entire network. The result of this
simulation is:
{\scriptsize
{\tt
\begin{verbatim}
1,0,0,0,0
0,1,0,0,0
0,0,1,0,0
0,0,0,1,0
0,0,0,0,1
1,0,0,0,0
0,1,0,0,0
\end{verbatim}
}
}
Each line in the output shows the overall states of nodes in the network. \texttt{1} denotes a node has
privilege. In the absence of 
state perturbation the invariant holds in this run. 
To inject faults with data poisoning,
the designer only needs to import the data poisoning library and
replace one statement in Figure \ref{code:sss}:  
{\scriptsize
{\tt
\begin{verbatim}
self.status =  Poison(0, size=1, rate=1,cascade=True)
\end{verbatim}
}
}
The \texttt{Poison} class implements data poisoning. 

\begin{figure}[t]
{\scriptsize
\begin{lstlisting}
NetworkSize = 5
SIMULATIONROUNDS=10
for i in range(NetworkSize):
    n = Node()
    Network.add(n)
for i in range(SIMULATIONROUNDS):
    for nid in range(NetworkSize):
        Network.nodes[nid].update()
\end{lstlisting}
}
\caption{Sample simulator} 
\label{code:simulator}
\end{figure}

\section{Different types of data poisoning.}
The architecture of data poisoning facilitates fine-grained control 
of fault injection. In this secton, we describe  three controllable 
dimensions of data poisoning.  We use Hoare triplets $\{P\}S\{Q\}$ to
define the semantics of data poisoning, where $\{P\}$ denotes predicates true before execution of statement
$S$ and $\{Q\}$ denotes predicates true after execution of $S$.
$S$ denotes basic expressions, i.e. assignment  expressions, or
conditional expressions that are used in control-flow statements.

\textbf{Determinism in Effect of Poisoning.}
Poisoned data may either always deviate, or deviate with some
probability.

\begin{defin}[Deterministic  Effect Poisoning]
$\forall S:$ $uses(S,v) \wedge$ \poi{v} $\in P \implies $  $ dev(S,v) \in Q$.
\end{defin}

\begin{defin}[Intermittent Effect Poisoning]
$\forall S:$ $uses(S,v) \wedge$ \poi{v} $\wedge p$  $\in P \implies $  $ dev(S,v) \in Q$, 
where $p$ is a predicate with a random Boolean value. 
\end{defin}

\textbf{Lifetime of Poisoned Data.}
Poisoned data may either  always be poisoned or have a single (generalizable to finite counts) poisoning.

\begin{defin}[Always Poisoned Data]
$\forall S:$ \poi{v} $\in P \implies $ \poi{v} $\in Q$.
\end{defin}

\begin{defin}[Transiently Poisoned Data]
$\forall S:$ \poi{v} $\in P \wedge uses(S,v) \implies \neg $  \poi{v} $\in Q$.
\end{defin}

\textbf{Infectiousness of  Poisoned data.}
Poisoned data can be defined as infectuous, causing poisoning for
variables derived from its value, similar to tainting  \cite{taint2010,taintcheck}. 

\begin{defin}[Infectious poisoning]
For an assignment expression, \texttt{x = u binOp v}:  \poi{u} $\in P \vee$ \poi{v} $\in P$ $ \implies $ \poi{x} $\in Q$, where
\texttt{binOp} is a binary operation. 
\end{defin}

\begin{defin}[Non-infectious poisoning]
For an assignment expression, \texttt{x = u binOp v}:  \poi{u} $\in P \vee$ \poi{v} $\in P \neg\implies  $\poi{x} $\in Q$, where
\texttt{binOp} is a binary operation.
\end{defin}

\begin{figure}
\begin{lstlisting}
if self.nid== 0:
   self.status =  Poison(0, infection_size=1, infection_rate=1,cascading=False)
else:
   self.status = 0
\end{lstlisting}
\caption{Non-infectious Transiently }
\end{figure}

\bibliographystyle{splncs03}
\bibliography{bibliography}

\end{document}